\documentclass[prb,twocolumn,showpacs,preprintnumbers,amsmath,amssymb,superscriptaddress]{revtex4}

\usepackage{graphicx}
\usepackage{dcolumn}

\newcommand{\beq}{\begin{equation}}
\newcommand{\eeq}{\end{equation}}
\newcommand{\beqa}{\begin{eqnarray}}
\newcommand{\eeqa}{\end{eqnarray}}

\newcommand{\kvec}{{\bf k}}
\newcommand{\qvec}{{\bf q}}
\newcommand{\Qvec}{{\bf Q}}

\begin{document}
\title{Fermi surface dichotomy in systems with fluctuating order}
\author{M. Grilli}
\affiliation{Dipartimento di Fisica,
Universit\`a di Roma ``La Sapienza", P.le Aldo Moro 5, I-00185 Roma, Italy}
\affiliation{SMC-INFM-CNR, Universit\`a di Roma ``La Sapienza", P.le Aldo Moro 5, I-00185 Roma, Italy} 
\author{G. Seibold}
\affiliation{Institut f\"ur Physik, BTU Cottbus, PBox 101344, 03013 Cottbus, Germany}
\author{A. Di Ciolo}
\affiliation{Dipartimento di Fisica,
Universit\`a di Roma ``La Sapienza", P.le Aldo Moro 5, I-00185 Roma, Italy}
\affiliation{SMC-INFM-CNR, Universit\`a di Roma ``La Sapienza", P.le Aldo Moro 5, I-00185 Roma, Italy} 
\author{J. Lorenzana}
\affiliation{SMC-INFM-CNR, Universit\`a di Roma ``La Sapienza", P.le Aldo Moro 5, I-00185 Roma, Italy} 
\affiliation{ISC-CNR, Via dei Taurini 19, I-00185, Roma, Italy} 
\begin{abstract}
We investigate the effect of a dynamical collective mode coupled with quasiparticles at specific
wavevectors only. This coupling describes the incipient tendency to
order and produces shadow spectral features at high energies, while
leaving essentially untouched the 
low energy quasiparticles. This allows to interpret seemingly contradictory 
experiments on underdoped cuprates, where many converging evidences 
indicate the presence of charge (stripe or checkerboard) order, which
remains instead elusive in the 
Fermi surface obtained from angle-resolved photoemission experiments. 
\end{abstract}
\date{\today}

\pacs{74.72.-h, 74.25.Jb, 71.18.+y, 71.45.-d}
\maketitle

\section{Introduction}
Strongly correlated systems like the heavy fermions and the
superconducting cuprates display excitations over 
a variety of energy scales. On short timescales (high energies)
electrons are excited incoherently over energies ranging 
from the highest local (Hubbard $U$) repulsion to the magnetic
superexchange interaction. On the other hand, 
over long timescales, i.e. at low energies around the Fermi level, the
excitations can acquire a coherent 
character typical of the long-lived Fermi-liquid quasiparticles (QPs).
These energy scales appear very clearly both in theoretical\cite{geo96} and
experimental\cite{ima98}  studies of the one particle spectral
function of strongly correlated systems. It is usually assumed that
the incoherent part has no momentum structure, an assumption which is  
emphasized by infinite dimensional studies where the self-energy
is momentum independent and spatial informations on quasi long range
correlations close to a phase transition are lost.\cite{geo96}

In this paper we discuss how this picture is modified in physical
spatial dimensions. We want to address how the spectral function 
looks like when the system is close to an ordered phase. 
This issue is particularly important  in the context of cuprates
where it has been proposed that some kind of stripe-like order
fluctuates in the metallic phase.\cite{zaa96b,cas98b,kiv98}

The scenario that we propose is based on the following qualitative
argument: In physical
 dimensions a system may have long (but finite) ranged order
parameter spatial correlations which are also long lived close to a 
quantum critical point. This defines a fluctuating frequency
$\omega_0$ above which the systems appears to be ordered. We argue 
that for energies larger than $\omega_0$ with respect to the Fermi
level the spectra should resemble the spectral function of an
ordered system. This spectral weight resides in what is usually called 
the incoherent part, which we argue, can have some important momentum
 structure. 
On the other hand at lower energies electrons average
over the order parameter fluctuations and ``sense'' a disordered
system. In this limit we expect Fermi-liquid QPs 
with all their well known characteristics like a Luttinger Fermi
surface (FS). 

To understand the momentum structure  of the
spectral function at energies higher than $\omega_0$
is important because if the incoherent part carries a memory of 
the close-by  ordered phase it should be possible to analyze it to 
obtain informations on what is the underlying fluctuating order.  
Usually ordered systems are well described by mean-field thus one can
obtain a first guess of how the incoherent part of the spectral 
function in the disordered phase should look like by performing a 
mean-field computation assuming long-range order. Comparison with
experimental data in the absence of long range order can be useful 
to identify the fluctuating order parameter. 

To fix ideas consider as an example a moderately large $U$ Hubbard system in a
half-filled bipartite lattice in two dimensions at $T=0$. In this case
an antiferromagnetic state is expected to be a competitive low-energy
state.  When the system is in the ordered
phase the spectral function will be reasonably well described by a mean-field 
computation and will show two Hubbard bands separated in energy by $m U$ with 
$m$ the staggered magnetization. The bands will show some dispersion 
governed by the scattering of the electrons with the mean-field
potential. Suppose that due to some frustrating effect long range
magnetic order is lost while keeping well formed magnetic moments. 
We expect that beyond mean field if $U$ is not too large (so that 
the disordered phase is metallic)
a QP will appear at the center of the Hubbard bands 
with small spectral weight resembling the dynamical mean field theory (DMFT)
picture.\cite{geo96} 
At high energies, however, electrons will sense a mean-field-like 
staggered potential for distances of the order of the correlation
length $\xi$, which can be quite long, and therefore the system will keep
substantial  memory of the mean-field like bands with their dispersion. 
Roughly we expect that the spectral function will look like the 
superposition of a Fermi-liquid-like spectral function, with a small
weight $z$ close to the Fermi level,
plus a blurred mean-field-like spectral function in the presence of
long-range order with a large weight $1-z$.  This is at first sight
similar to the DMFT picture but it differs in that in 
DMFT there are no magnetic correlations surviving in the
disordered phase and the incoherent part becomes momentum 
independent. We will show that in finite dimensions the incoherent part 
carries important informations encoded in the momentum dependence
which within a DMFT approach would require the cluster extensions
developed more recently.~\cite{maier05}
 
\begin{figure}[hbt]
\includegraphics[angle=180,width=6cm,clip=true]{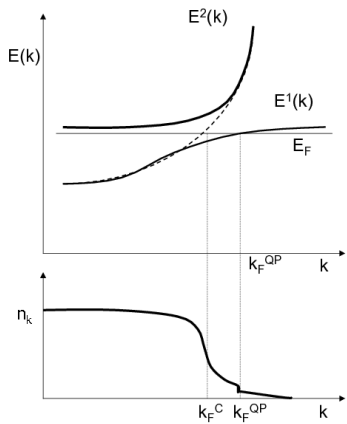}
\caption{Schematic view of heavy-fermion system with a Kondo-like resonance arising at the Fermi energy E$_F$
from the mixing of a deep narrow $f$ level (not shown) and the conduction band (dashed line). Two QP bands arise
$E^{1,2}(k)$. The corresponding momentum distribution function $n_k$ is shown below with a true Fermi momentum
$k_F^{QP}$ and a ``fictitious'' Fermi surface at $k_F^c$.
}
\label{Fig1-DFS}
\end{figure}

Another example can clarify the concept of an incoherent part with a
strong momentum dependence which carries  physical information on
the short range physics. Lets consider the more standard issue of
large and small FSs in heavy fermions represented in
Fig.~\ref{Fig1-DFS}. 
In heavy fermions strongly correlated electrons in a  narrow
half-filled $f$ level hybridize with 
electrons in a conduction band and give rise to a Kondo resonance at the Fermi level formed by coherent
QP states. The width (and weight) of this QP band is usually quite small and sets the scale of the coherence
energy in these systems. Now consider the momentum distribution
function defined by:
\beq
n_\kvec \equiv \int d\omega A(\kvec, \omega) f(\omega),
\label{nk}
\eeq
where $f(\omega)$ is the Fermi function and the spectral density 
\beq 
A(\kvec,\omega)\equiv\frac{1}{\pi}ImG(\kvec,\omega)=\frac{1}{\pi}
\frac{\Sigma''(\kvec,\omega)}{\left(\omega-\Sigma'(\kvec,\omega) \right)^2+\Sigma''^2(\kvec,\omega)}
\eeq
is proportional to the imaginary part of the electron Green function with real (imaginary) part
of the self-energy $\Sigma'$ $(\Sigma'')$.
It is crucial to recognize that $n_\kvec$ involves all the excitation energies and its features
might be dominated by the incoherent part of the spectrum if the QPs
have a minor weight. Indeed strictly speaking the true FS at zero temperature is given by the small jump
in the Fermi distribution function determining the Fermi momenta of the QPs at $\kvec_F^{QP}$. 
This FS is large and satisfies the Luttinger theorem with a number of carriers including the
electrons in the $f$ level. This FS would naturally be determined by following the QP dispersion. On the other hand
the shape of $n_\kvec$ is substantially determined by the (incoherent)
part of the spectral function, which
has strong weight at energies corresponding to both the $f$ level and  the conduction band. This latter
gives rise to a rather sharp decrease of $n_\kvec$ at a ``fictitious''
Fermi momentum $\kvec^c$, corresponding 
to the FS that the electrons in the conduction band would have in the
absence of mixing with the f-level. If the hybridization between the
$f$ level and the conduction band is turned off so that the
QP weight $z$ is driven to zero one  reaches
a situation in which the decrease at $\kvec^c$ becomes a discontinuity 
and the small jump at $\kvec_F^{QP}$ disappears.   It is clear that the 
sharp decrease of spectral weight
at $\kvec^c$ for finite hybridization has strong physical content for
an observer who ignores the underlying model.  Mutatis mutandis this shows that a
computation in which the fluctuating order is artificially frozen can give
some hints on the distribution of the incoherent spectral weight in
the less trivial case with fluctuations.

The above ideas  but with a more complicated order parameter may explain
perplexing data on cuprates.  Several years ago underdoped
La$_{2-x}$Sr$_x$CuO$_4$ (LSCO) compounds
were examined and a dichotomy was found in the Fermi
surface (FS) determined by two different 
treatments of the data. On the one hand the momentum dependence of the
low-energy part of the energy 
distribution curves was followed, thereby reconstructing the low-energy 
QP dispersion. In this way a large FS was found corresponding to the
Fermi-liquid LDA band-structure and fulfilling the Luttinger 
requirement that the volume of the FS encircled the whole number of
fermionic carriers $n=1-x$. 
On the other hand the FS was determined from the momentum distribution
$n_\kvec$, obtained by integrating 
the spectral function over a broad energy window ($\sim 300$
meV). Then the locus of momentum-space  
points where $n_\kvec$ displays a sharp decrease, marked a FS formed 
by two nearly parallel (weakly
modulated) lines along the $k_x$ direction and crossing two similar
lines along the $k_y$ direction. 
This crossed FS would naturally arise in a system with
one-dimensional stripes along the $x$ and $y$ 
directions. 

As a matter of fact the stripe Fermi surface has been
observed both in systems which show a striped ground state
and in systems where long-range stripe order has not been
detected.\cite{zho99,zho01} As in the heavy Fermion case for
relatively small  $z$  we expect that 
$n_\kvec$ is dominated by the
incoherent spectral weight. According to our scenario
$n_\kvec$ should resemble the Fermi surface of 
stripes in mean-field regardless of whether stripe order is static or
 fluctuating at low frequencies. 
Indeed LDA computations in the presence of stripes with
magnetic and charge long-range order reproduce this Fermi surface.\cite{ani04}

The experimental stripe-like Fermi surface is not flat as could be
expected for a 
perfect one dimensional band structure but shows some wavy
features, which depend on the details of residual hopping processes 
perpendicular to the stripes. Remarkably  the wavy features are well
reproduced 
by the LDA computation both with regard to amplitude and periodicity, giving
credibility to the idea that the static LDA computation provides a snapshot 
of the fluctuating order in the disordered phase (cf. also Ref. 
\onlinecite{gra02} for a more phenomenological approach).

The main point is that the formation of QPs due to some
coherence effect is a small perturbation for the overall
distribution of spectral weight. Thus a careful study of $n_\kvec$ can
give precious information on the proximity to some  ordered phase. 
By the same token 
the presence of QPs and a Fermi-liquid-like Fermi surface
are not incompatible with two-particle responses (neutrons, optical) 
which show strong features of ordering (like stripes) at frequencies
above $\omega_0$. This explains why 
computations of fluctuations on top of stripe phases  
with long range order\cite{sei05,lor03} explain well optical
conductivity and neutron scattering data of systems without long-range
order.  It also explains how nodal quasiparticles can coexist with
fluctuating stripe order. 

In the following section we present a toy model 
of the  Kampf-Schrieffer-type\cite{kam90} where the dichotomy between
a Fermi-liquid-like Fermi surface and a momentum dependent incoherent
part, reflecting the  fluctuating order,  
can be illustrated. Although we study self-energy
corrections self-consistently, the lack of vertex corrections
makes our computations  reliable only in weak coupling where QP
weights are close to one. Within this limitation  one can show 
that the described scenario  holds.

In Sec.~\ref{sec:charge-coll-mode} we present the model,
while in Sec.~\ref{sec:numer-results-fluct} we present some numerical
results. A discussion of the results and our 
conclusions are reported in Sec.~\ref{sec:disc-concl}.

\section{The  model}
\label{sec:charge-coll-mode}
In order to substantiate the above ideas
we consider a system of electrons coupled to a dynamical order
parameter which can describe charge ordering
(CO) fluctuations or spin ordering (SO) fluctuations or both. To fix ideas we consider
CO fluctuations  described by an effective action
\begin{equation}
S=-g^2 \sum_{\bf q}  \int_0^{\beta}d\tau_1\int_0^{\beta}d\tau_2
\chi_{\bf q}(\tau_1-\tau_2) \rho_{\bf q}(\tau_1) \rho_{-{\bf q}}(\tau_2).
\end{equation}

In order to simplify the calculations we consider a Kampf-Schrieffer-type
model susceptibility \cite{kam90}
which is factorized into an $\omega$- and q-dependent part, i.e.
\begin{equation}\label{chi}
\chi_{\bf q}(i\omega)=W(i\omega)J({\bf q})
\end{equation}
The (real-frequency)-dependent part is
$W(\omega)=\int d\nu F_{\omega_0,\Gamma}(\nu) 2\nu/(\omega^2-\nu^2)$, with $F_{\omega_0,\Gamma}$ being
a normalized lorentzian distribution function centered around $\omega_0$ with 
half-width $\Gamma$, 
$F_{\omega_0,\Gamma}(\omega)\sim \Gamma/[(\omega-\omega_0)^2 +\Gamma^2]$.
The momentum-dependent part in $D$ dimensions reads
\begin{equation}\label{jq}
J_D({\bf q})={\cal N}
\sum_{\eta=1}^D \sum\limits_{\pm Q_\eta}\frac{\gamma}{\gamma^2
+1-\cos(q_\eta-Q_\eta)}.
\end{equation}
${\cal N}$ is a suitable normalization factor introduced to
keep the total scattering strength constant while varying $\gamma\propto \xi^{-1}$
(with $\xi$ the CO/SO correlation length).
To simplify the treatment and to make the  effect of fluctuations as clear as possible, we will 
mostly consider the case of spatially coherent fluctuations. Although
formally the correlation length $\xi$ diverges, the ordering is not
static and we will show that this is enough for the spectral function
to converge to the Fermi-liquid Fermi surface at zero frequency. 

The infinite correlation length case is described in $D$ dimensions by,
\begin{equation}\label{jqdelta}
J_D({\bf q})=\frac{ 1}{4}
\sum_{\eta=1}^D 
\delta(q_\eta -Q_\eta)+ \delta(q_\eta +Q_\eta).
\end{equation}
In previous works the spatially-smeared version of 
$J_D(\qvec)$, Eq. (\ref{jq}), 
was considered to describe the kink in the electron dispersions \cite{sei01b}
and the (still experimentally controversial \cite{gwe04,DESSAU}) isotopic 
dependence of these dispersions.~\cite{sei05b}  The susceptibility $\chi$
contains the charge-charge correlations in the case of charge
fluctuations and spin-spin correlations in the case of magnetic
fluctuations. 

If only the dynamical part $W(\omega)$ were present in $\chi_\qvec(i\omega)$,
one would have a bosonic spectrum
$B(\omega)= \mbox{tanh}(\omega/(kT)) F_{\omega_0,\Gamma}(\omega)$
which is a ``smeared'' version of the Holstein phonon considered
in Ref. \onlinecite{eng63}. The crucial feature of the susceptibility
(\ref{chi}) is the substantial momentum dependence, which describes
the (local) order formation and reflects the proximity to an
instability with broken translational symmetry.

The static limit $F_{\omega_0,\Gamma}(\nu)=\delta(\nu)$  together with an 
infinite charge-charge correlation length ($\gamma \rightarrow 0$) as the
one considered in Eq.~(\ref{jqdelta})
reproduces mean-field results for a long-range phase.~\cite{kam90,sei00}
The static limit of (\ref{chi}) has been the object of an intense activity
based on the idea, pioneered in Ref.~\onlinecite{lee73}, that different types of 
slow order-parameter (OP) fluctuations (SO \cite{kam90} or CO \cite{sei00})
can be treated as classical fluctuations with 
time-independent correlations. The equivalence of such static degrees
of freedoms 
with quenched impurities has more recently been formalized  
and cast in field-theoretical language.~\cite{pos03} Although they allow for
(nearly) exact solutions, the main drawback of these static approaches is that
they do not allow for the aimed separation of energy (i.e. time)
scales since to 
justify the static character of the OP fluctuations, they
assume that their relaxation time  $\tau_{OP}$ is much longer 
than the inelastic scattering rate of the electrons 
$\tau_e$, $\tau_{OP}\gg \tau_e$.~\cite{pos03} Here we consider instead 
\beq
F_{\omega_0,\Gamma}(\nu)=\delta(\nu-\omega_0)
\eeq
representing a $dynamical$ fluctuation oscillating at a frequency $\omega_0$ and therefore 
averaging out on timescales larger than $\tau_{OP}\sim 1/\omega_0$.

In the present limit the problem  has also a simple Hamiltonian
formulation which we introduce for later use:
\begin{eqnarray}
  \label{eq:ham}
&&H=\sum_{\kvec\sigma} \xi_{\kvec} c_{\kvec\sigma}^\dagger
c_{\kvec\sigma}+ \omega_0 (b_{\Qvec}^\dagger b_{\Qvec} 
+b_{-\Qvec}^\dagger b_{-\Qvec})\\
&&+ g \sum_{\kvec\sigma} \left[c_{\kvec-\Qvec \sigma}^\dagger c_{\kvec\sigma}
(b_{\Qvec}^\dagger+ b_{-\Qvec})+ c_{\kvec+\Qvec \sigma}^\dagger c_{\kvec\sigma} (b_{-\Qvec}^\dagger+ b_{\Qvec}) \right]  \nonumber
\end{eqnarray}
where  $c_{\kvec\sigma}^\dagger$ creates a free Fermion and $b_{\pm\Qvec}^\dagger$
creates a bosonic collective mode excitation and
$\xi_{\kvec}\equiv\varepsilon_{\kvec}-\mu$ with $\varepsilon_{\kvec}$
the non interacting dispersion relation and $\mu$ the chemical
potential.

QPs at energies much larger than $\omega_0$ see an essentially static fluctuation and 
modify their dispersion as in the mean-field calculation mentioned above. For them
the fluctuations are static and, if the momentum dependent
part $J(\qvec)$ is strongly peaked around the ordering wavevector $\Qvec$,
they are scattered like in the presence of a long-range symmetry breaking. Their dispersions are 
modified accordingly. On the other hand, for QPs at energies in a shell of width $\omega_0$
around the Fermi level the order parameter fluctuations are far from being static and 
have not enough energy to excite them.
For these QPs the system is still in a uniform Fermi-liquid metallic state.
Therefore, if their dispersion is followed near the chemical potential
$\mu$, a large FS fully 
preserving the Luttinger volume is found. 
In the following we explicitly calculate the band dispersions and the FS 
to show this physical mechanism at work and the resulting dichotomy of the FS.

Neglecting vertex corrections we iteratively solve at zero temperature
the following   
coupled set of equations for 
the self-energy and the Green's function   
\begin{eqnarray} 
\Sigma(\kvec,i\omega)&=&-\frac{g^2}{\beta} 
\sum_{\qvec,ip} \chi_\qvec(ip) G(\kvec-\qvec,i\omega-ip) \label{SELF}\\ 
G(\kvec,i\omega)&=&\frac{1}{i\omega-\xi_\kvec-\Sigma(\kvec,i\omega)} \label{GF}. 
\end{eqnarray} 
To the best of our knowledge our self-consistent treatment goes beyond
previous solutions that 
remained at the perturbative level. The latter can be obtained
 at lowest order by replacing the Green's function in Eq.~(\ref{SELF}) 
by the non-interacting $G^0$. For our specific model of 
the susceptibility the self-energy becomes: 
\begin{equation} 
\Sigma^{(1)}(\kvec,i\omega)= g^2\left\lbrace \frac{f(\xi_{\kvec+\Qvec})}{i\omega +\omega_0 
-\xi_{\kvec+\Qvec}}+\frac{1-f(\xi_{\kvec+\Qvec})}{i\omega -\omega_0 
-\xi_{\kvec+\Qvec}}\right\rbrace \label{self0} 
\end{equation} 
where the superscript $(1)$ denotes the order of the iteration  
and $f(\xi_{\kvec+\Qvec})=\Theta(-\xi_{\kvec+\Qvec})$  
is the occupation number. 
Inserting this self-energy into Eq. (\ref{GF}) leads to a Green's function  
$G^{(1)}(\kvec,i\omega)$ which displays now two poles and which can be used 
to compute $\Sigma^{(2)}(\kvec+\Qvec,i\omega)$ (and thus $G^{(2)}(\kvec+\Qvec,i\omega)$)  
and so on. 
The detailed procedure is described in the Appendix.

For the sake of  simplicity, we first implement our numerical analysis in one dimension
and then we describe the case of two dimensions, which is more
relevant for layered materials.

\section{Numerical results of fluctuating order}
\label{sec:numer-results-fluct}
\subsection{Numerical analysis in one dimension}
\subsubsection*{Coherent order parameter fluctuations}
\begin{figure}[hbt]
\includegraphics[width=9cm,clip=true]{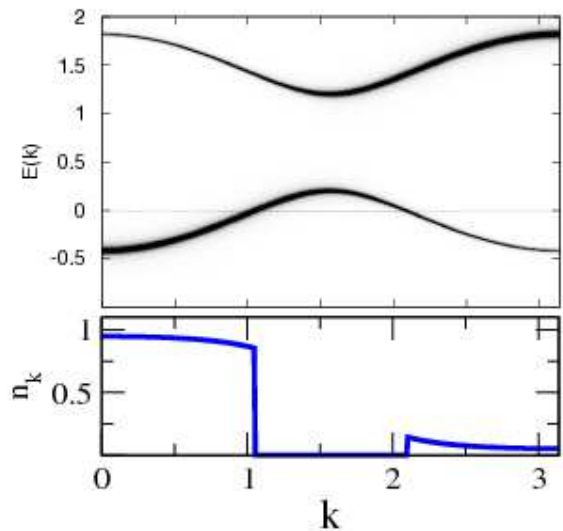}
\caption{(Color online) Upper panel: Two-pole band structure 
in one dimension for
$g=0.5t$, $\omega_0=0$ and particle density $n=0.67$. 
The width of the curve is proportional to the weight of 
the state and energies are measured with respect to E$_F$. 
Lower panel: Momentum distribution curve.}
\label{Fig2-DFS}
\end{figure}

We first present the results for a one-dimensional model with a commensurate
ordering wavevector $Q=\pi$ corresponding to CO/SO 
with a doubling of the unit cell. 
In particular we  consider a band of QPs
$
\varepsilon_\kvec=-\cos(\kvec) 
$
(we choose a unit hopping parameter $t=1$ and a unitary lattice spacing) 
coupled via a CO/SO mode given by Eqs. (\ref{chi}) and (\ref{jqdelta}).
Fig.~\ref{Fig2-DFS} reports the simplest case in which  
only two poles are 
considered in the
Green's function given by $G^{(1)}$
for a generic filling ($n=0.67$: here and in the following densities are
defined as total number of particles per site). 
%
This is the standard mean-field solution, with a doubling of the unit
cell and a folding of 
the bands resulting in a double FS. In the absence of interaction the
Fermi momentum would be given by $k_F=n\pi/2=1.05$. The dispersion
relation instead displays two Fermi points and a very different Fermi
surface volume, ``violating'' the Luttinger theorem. 

\begin{figure}[hbt]
\includegraphics[width=9cm,clip=true]{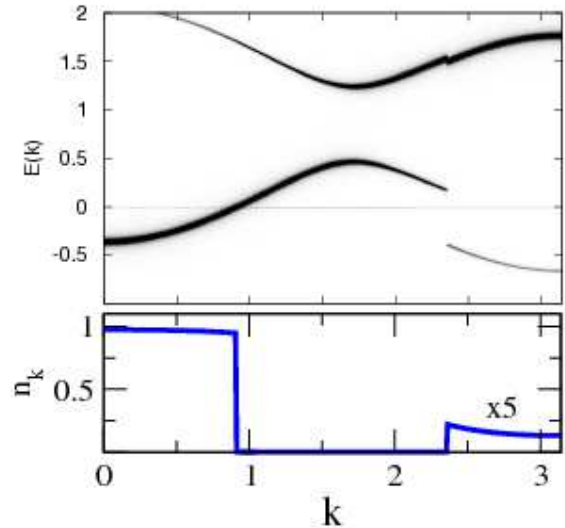}
\caption{(Color online) Upper panel: Two-pole band structure in one dimension for $g^2/\omega_0=0.5$, $\omega_0=0.3$ and
 electron density $n= 0.57$. The width of the curve is proportional to the 
weight of the state and energies are measured with respect to E$_F$. 
Lower panel: Momentum distribution curve. For $k>2$ the occupation number 
$n_k$  has been scaled by a factor $5$ to enhance the visibility.
}
\label{Fig3-DFS}
\end{figure}

The situation is very different for a {\it dynamical} fluctuation.
In Fig.~\ref{Fig3-DFS} we report the spectral function 
 with $\omega_0=0.3$ and moderately weak coupling
$g^2/\omega_0=0.5$. Also in this case shadow bands appear 
so that the electronic structure has bands 
close to the location of the  bands in the broken symmetry state. 
The shadow bands, however, do not reach the Fermi level and 
one has only one FS point at $k=k_F$. Therefore
the Luttinger theorem is satisfied since the divergent pole
in $G(k_F^2,\omega=0)$ from the static solution now turns into
a step so that $G(k,\omega=0)<0$ for $k>k_F$.

\begin{figure}[hbt]
\includegraphics[width=9cm,clip=true]{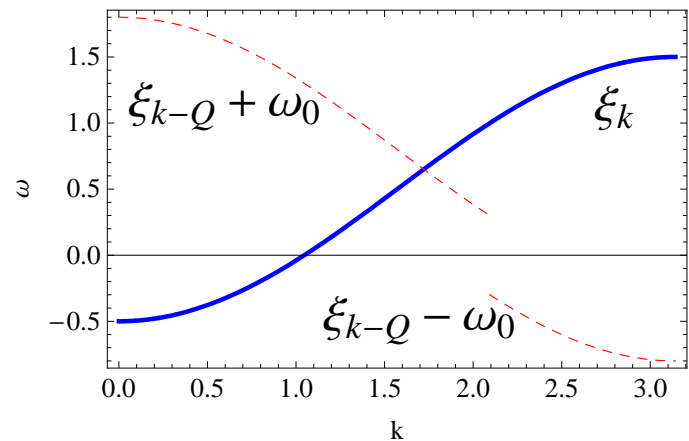}
\caption{(Color online) Schematic electronic structure for $g=0$.
}
\label{Fig4-DFS}
\end{figure}

The spectral function can be written in the Lehmann  
representation on the real axis as:  
\begin{eqnarray}
  \label{eq:lehman}
A(\kvec,\omega>\mu)&=&\sum_{\nu} |\langle \phi^{N+1}_\nu | c_{\kvec\sigma}^\dagger |
\phi^N_0  \rangle|^2 \delta(\omega-E^{N+1}_\nu+E^N_0)\nonumber\\
&&\\
A(\kvec,\omega<\mu)&=&\sum_{\nu} |\langle \phi^{N-1}_\nu | c_{\kvec\sigma} |
\phi^N_0  \rangle|^2 \delta(\omega-E^N_0+E^{N-1}_\nu)\nonumber
\end{eqnarray}
$A(\kvec,\omega)$ has structure at the energies of the excitation
of the system with one added particle ($\omega> \mu$) or one removed
particle ($\omega< \mu$). The result of Fig.~\ref{Fig3-DFS}
can be understood from the low energy excitation  of the system when
$g=0$. These are listed in Table~\ref{tab:excit} and the corresponding
excitation energies are shown in Fig.~\ref{Fig4-DFS}. For $g=0$
all the weight is in the main band labeled $\xi_\kvec$ because the
matrix elements in Eq.~\eqref{eq:lehman} vanish when one boson is
present. The effect of a finite $g$ is to give some spectral weight to
the ``shadow'' band at $\xi_{\kvec-\Qvec}\pm \omega_0$ and to introduce some
level repulsion when the bands cross. The important point is that the
shadow band never touches the Fermi level but it is separated from it by the
energy to create the bosonic excitation. We associate the main band
with the QP band and the shadow band with the incoherent
spectral weight. Clearly  close enough to the
Fermi level only the QP band  exists.

\begin{table}[htbp]
\begin{ruledtabular}
  \begin{tabular}{ccc}
\multicolumn{3}{c}{Addition $\omega > 0$}\\
 State                                 & $E^{N+1}_\nu-E^N_0$ & Momentum   \\
 $c_{\kvec\sigma}^\dagger |\phi^N_0\rangle$   &$\xi_\kvec$ & $\kvec$\\
$c_{\kvec-\Qvec\sigma}^\dagger b_{\Qvec}^\dagger |\phi^N_0\rangle$
&$\xi_{\kvec-\Qvec}+\omega_0$ & $\kvec$ \\
\hline
\multicolumn{3}{c}{Removal $\omega < 0$}\\
 State                                 & $E^N_0-E^{N-1}_\nu$ & Momentum   \\
 $c_{-\kvec\sigma} |\phi^N_0\rangle$   &$\xi_{-\kvec}$ & $\kvec$\\
$c_{-\kvec+\Qvec\sigma} b_{-\Qvec}^\dagger |\phi^N_0\rangle$
&$\xi_{-\kvec+\Qvec}-\omega_0$ & $\kvec$ 
  \end{tabular}
\end{ruledtabular}
  \caption{Low energy excitations in the $g=0$ limit. The central
 column shows the excitation energy. Notice that $\xi_{-\kvec}=\xi_{\kvec}$.}
  \label{tab:excit}
\end{table}

Notice that the shadow bands are quite similar to the case of a
symmetry-broken state 
(but for a shift in energy of order $\omega_0=0.3$, which is small on the
scale of the hopping $t=1$). 

Because the QP states at the Fermi level are negligibly affected by the scattering with the dynamical ordering mode,
the FS (here represented by points) is preserved and no
shadow branches appear at low energy. Therefore already this simple weak-coupling case 
shows that an ordered state would be inferred from the presence of shadow bands at high energies, while
the low-energy QP are characteristic of a uniform state.

In Fig.~\ref{Fig3-DFS} we also report the momentum distribution
$n_\kvec$. From this quantity one can see that some weight is indeed present at high energy via 
the shadow band appearing at
momenta above $k^*\approx 2.3$. This is a signature of some ordering in the system. However,
since the CO/SO is dynamical  the dispersion discontinuously jumps from above to below
the Fermi level without producing an additional branch to the FS and with no violation of the Luttinger theorem.
In this last respect we explicitly checked that the Green function at zero frequency
is positive for $k<k_F$, it changes sign through a pole at $k_F$ and stays negative all the way
up to $\pi$. The Fermi momentum $k_F$ is the same that one would have in a non-interacting
tight-binding system with the same number of particles $k_F=n \pi/2$ with $n=0.57$ in this case.

\begin{figure}[hbt]
\includegraphics[width=9cm,clip=true]{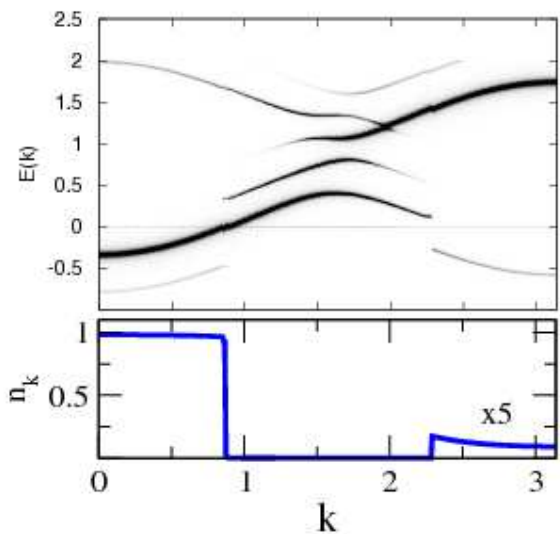}
\caption{(Color online) Upper panel: Six-pole band structure in one 
dimension for $g^2/\omega_0=0.5$ and $\omega_0=0.2$. 
The electron density is $n=0.55$. 
The width of the curve is proportional to the weight of 
the state and energies are measured with respect to E$_F$. 
Lower panel: Momentum distribution curve.
For $k>2$ the occupation number 
$n_k$  has been scaled by a factor $5$ to enhance the visibility.
}
\label{Fig5-DFS}
\end{figure}

The above picture remains valid if additional poles are considered in the Green function
provided the QP-CO/SO-mode coupling is in a weak-to-moderate regime. Fig.~\ref{Fig5-DFS}
reports the case of three poles for the same parameters of Fig.~\ref{Fig3-DFS}.
The increased allowed number of poles better represents the shift of spectral weight at
high energies induced by multiple shadow bands. Nevertheless, the finite frequency of the
mode protects the low-energy QPs from being scattered and leaves the low-energy spectrum
unaffected: The FS is still formed by just two points despite the appearance of several
shadow bands at high energy typical of a CO/SO state.  The effect of
the self consistency is to ``blur'' the shadow bands but still the incoherent
spectral weight retains a strong momentum dependence. 

\begin{figure}[hbt]
\includegraphics[width=9cm,clip=true]{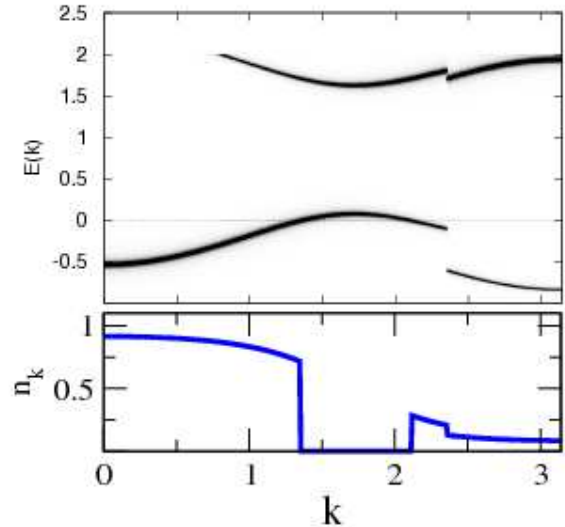}
\caption{(Color online) Two-pole band structure in one dimension 
for $g^2/\omega_0=2.$ and $\omega_0=0.3$. 
The electron density is $n=0.83$. The width of the curve is proportional to the weight of 
the state and energies are measured with respect to E$_F$. 
Lower panel: Momentum distribution curve.
}
\label{Fig6-DFS}
\end{figure}

Upon increasing the QP-CO/SO mode coupling $g$ one eventually enters a regime where
our non-crossing perturbative scheme neglecting vertex corrections breaks down and
the Luttinger theorem is violated. This situation is reported in Fig. \ref{Fig6-DFS},
where a second branch of the FS appears due to the strong bending down of the band at
momenta $k\approx k^*$. In this case the Green function at zero frequency $G(\omega=0)$
changes sign twice (and diverges) at the two Fermi momenta $k_{F1}$ and $k_{F2}$ while
it changes sign passing through zero at $k=k^*$, where the jump in the dispersion
below $\mu$ signals a divergence of the self-energy (\ref{self0}). One can check
explicitly that the volume corresponding to a positive $G(\omega=0)$ 
is larger than the one given by occupied states in the non-interacting system thus
Luttinger theorem is not satisfied. We believe this is an
artifact due to the lack of vertex corrections which become important
as the coupling is increased.

The apparent success of this simple theory on illustrating the
spectral function of a system with fluctuating order is encouraging. 
However, a different problem arises if one considers the half-filled
case.  Without long range order we would expect a metallic state if we
were able to solve the model exactly. On the contrary 
commensurate scattering in the perturbative solution 
produces the unphysical result of a gaped FS in contrast with
the fact that no true broken symmetry is present in the system.
 On the other hand it is not surprising that our perturbative approach
fails when our singular interaction $\chi(\qvec,\omega)$ connects two degenerate states at the FS
as it occurs in this case. We believe this failure is due to the lack
of vertex corrections which we expect to suppress the scattering
at the Fermi level. We have tested this idea phenomenologically  by
assuming that the QPs at the Fermi level are protected against this singular
scattering by a momentum dependence of the coupling of the form
\beq
\tilde g (\tilde\varepsilon_\kvec,\tilde\varepsilon_{\kvec+\qvec})=g \tanh 
(\frac{\tilde\varepsilon_{\kvec}}{\omega_0})
\tanh(\frac{\tilde\varepsilon_{\kvec+\qvec}}{\omega_0})
\label{renvertex}
\eeq
One technical remark is in order here. Perturbatively one could introduce on the r.h.s.
of Eq. (\ref{renvertex}) the bare QP dispersion $\varepsilon_\kvec$. However, 
at moderate-strong couplings the
QP dispersions are substantially modified by the coupling with the modes to 
dressed QP dispersions $\tilde \varepsilon_\kvec$. To suppress the
scattering of the {\it dressed}  QPs near the true FS,
one must insert in (\ref{renvertex}) the renormalized dispersions $\tilde \varepsilon_\kvec$.
This need of a self-consistency scheme considerably complicates the calculations
and led us to consider only simple symmetry breakings (cell doublings).

\begin{figure}[hbt]
\includegraphics[width=8cm,clip=true]{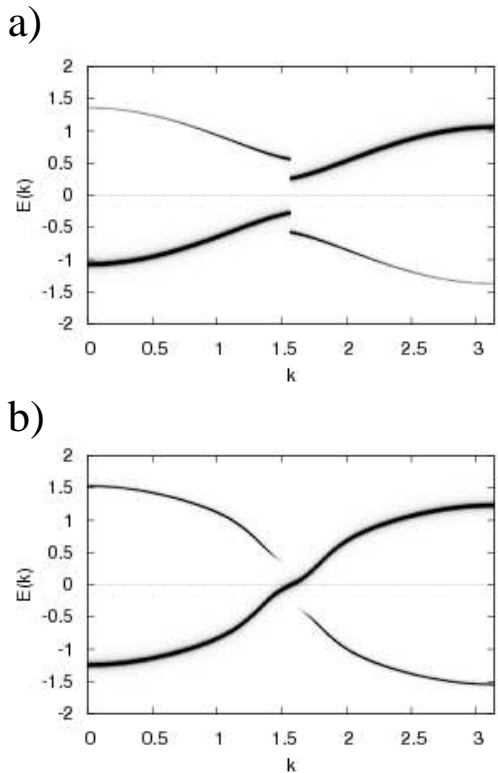}
\caption{Two-pole band structure in one dimension for $g^2/\omega_0=0.5$ and $\omega_0=0.3$
at half-filling $n=1$. In (a) the bare vertex $g$ is used, while in (b) the phenomenological 
vertex $\tilde g$ of Eq. (\ref{renvertex}) is used.
}
\label{Fig7-DFS}
\end{figure}

Although the form (\ref{renvertex}) is adopted  on a purely phenomenological basis, we like to remind that
several microscopic calculations \cite{gri94,san05} show that in strongly correlated systems
the coupling between QPs and phonons is severely suppressed at low energies. This suppression is 
mostly effective when the exchanged momenta $v_F \qvec$ are larger than the typical exchanged energy
$\omega_0$. This is the case here, where the exchanged momenta are peaked at sizable $\Qvec$'s.

Figs.~\ref{Fig7-DFS}(a) and (b) illustrate the effect of this vertex correction in the half-filled case $n=1$,
where it plays a crucial role to restore the metallicity of the system. For the previously 
considered cases of generic filling we find that the vertex corrections play a minor role and the results
obtained with $\tilde g$ differ little at moderate coupling from those reported above. However, for strong coupling the phenomenological 
form of the vertex Eq. (\ref{renvertex})
again prevents the system to violate the Luttinger  theorem as e.g. in case of the result
shown in Fig.~\ref{Fig6-DFS}.

\subsubsection*{Finite range fluctuations}
\begin{figure}[hbt]
\includegraphics[width=9cm,clip=true]{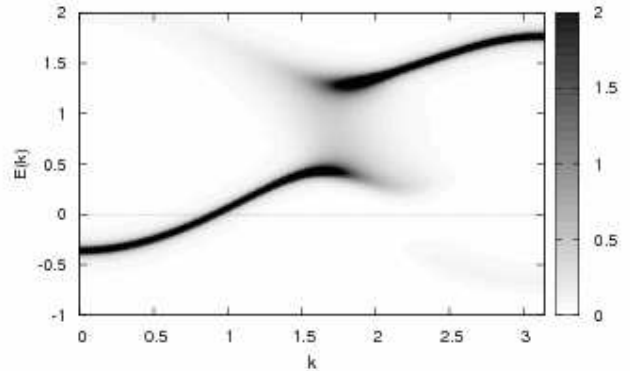}
\caption{Two-pole band structure in one dimension 
for $g^2/\omega_0=0.5$ and $\omega_0=0.3$
at generic filling $n=0.57$. 
Here a finite inverse coherence length $\gamma =0.02$ and time 
$\Gamma =0.1$ are used.}
\label{Fig8-DFS}
\end{figure}

Since in realistic systems the dynamical fluctuations have a finite coherence length
and a finite lifetime, we also investigated the case of a one-dimensional QP band exposed
to fluctuations having finite $\Gamma$ and $\gamma$ in $F_{\omega_0,\Gamma}$ and in Eq. (\ref{jq}).
Obviously the finite extension in space and time of the fluctuations
produces broadening in the spectra. As it is natural, one finds that the broadening in the momentum
direction is ruled by $\gamma$, while $\Gamma$ rules the broadening along the energy axis.
As it clearly appears in Fig.~\ref{Fig8-DFS}, the broadening of the spectra does not spoil the
essential feature of the coupling with a dynamical mode (cf. 
Fig.~\ref{Fig3-DFS} for $\gamma=\Gamma=0$ but same parameters otherwise): 
The shadow bands persist as high-energy signatures
of a (local) order, while the FS stays unchanged and Luttinger theorem is 
obeyed provided the mode is sufficiently
narrow in energy ($\gamma <\omega_0$) so that one can neglect the ``leakage'' of weight down to the Fermi level.

\subsubsection*{Pairing effects}
\begin{figure}[hbt]
\includegraphics[width=9cm,clip=true]{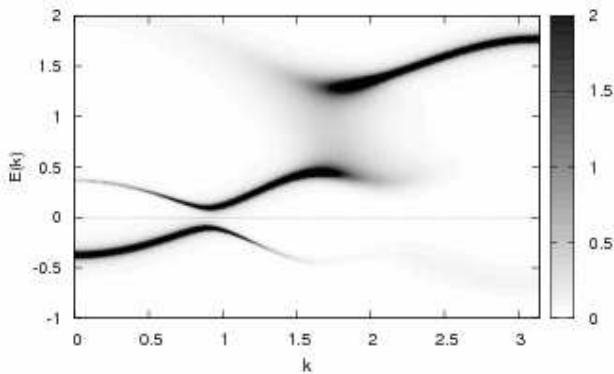}
\caption{Two-pole band structure in one dimension for $g^2/\omega_0=0.5$ and $\omega_0=0.3$
at generic filling $n=0.4$. Here a finite inverse coherence length 
$\gamma=0.02$ and time $\Gamma=0.1$ are used as
well as a finite pairing gap $\Delta=0.1$.}
\label{Fig9-DFS}
\end{figure}

We further explore our one-dimensional model to investigate the effects of a particle-particle pairing.
To this purpose we introduce in our bare QP band structure a finite gap $\Delta$, which, 
as it is usual 
for this type of pairing, is tight to the FS. Accordingly  the Bogoliubov particle-hole mixing occurs
near the FS and the QP band opens a gap centered
at the Fermi level. The branch below the Fermi level bends down giving rise to a maximum occurring at
a momentum coinciding with the Fermi momentum of the unpaired QPs.
Fig.~\ref{Fig9-DFS} displays this effect for our typical parameter set at a generic filling.
In particular one can see that the particle-particle pairing 
only modifies the low-energy states over scales of order $\Delta$, while 
leaving unchanged the high-energy
states, which still display the clear effect of the dynamical
ordering.
This finding shares close resemblance with recent scanning tunnelling experiments in cuprates
\cite{kohsaka}, where an energy scale $\Delta_0$ (analogous to our $\omega_0$) separates the
low-energy Bogoliubov coherent QPs from the high-energy excitations carrying information
of a charge ordering tendency. The one-dimensional character of our
toy model helps in separating 
these different excitations in momentum space. 
A more detailed analysis of this issue would require modeling specific
features of the cuprates and is beyond the scope of the present paper,
but we find  this preliminary one dimensional finding rather suggestive.

%

\subsection{Numerical analysis in two dimensions}
Once the main effects of the dynamical mode exchange have been presented in one dimension,
we illustrate the two-dimensional case. For the bare electron dispersion we use: 
\begin{displaymath}
\varepsilon_\kvec = -2t(\cos(k_x)+\cos(k_y)) - 4t' \cos(k_x)\cos(k_y) 
\end{displaymath}
For concreteness the parameters have been chosen to reproduce  the FS of LSCO, taking
$t=342$ meV, $t'/t=-0.2$, and $\mu=-0.1t$. For the mode frequency and coupling, we choose
$\omega_0=50$ meV and $g^2=1.5 \omega_0$, while we consider only coherent fluctuations
with vanishing broadenings ($\Gamma=\gamma=0$). To avoid too many band foldings, which could 
complicate the analysis, we choose a order given by a simple cell doubling in the $x$ direction.
This gives rise to $\Qvec=(\pm \pi, 0)$. Of course this does not correspond to the charge/spin
modulation observed in LSCO cuprates, but it better fits our simplicity and illustrative
purposes (on the other hand it corresponds to the spin ordering in
FeAs stoichiometric compounds\cite{don08,cla08}).

\begin{figure}[hbt]
\includegraphics[width=8cm,clip=true]{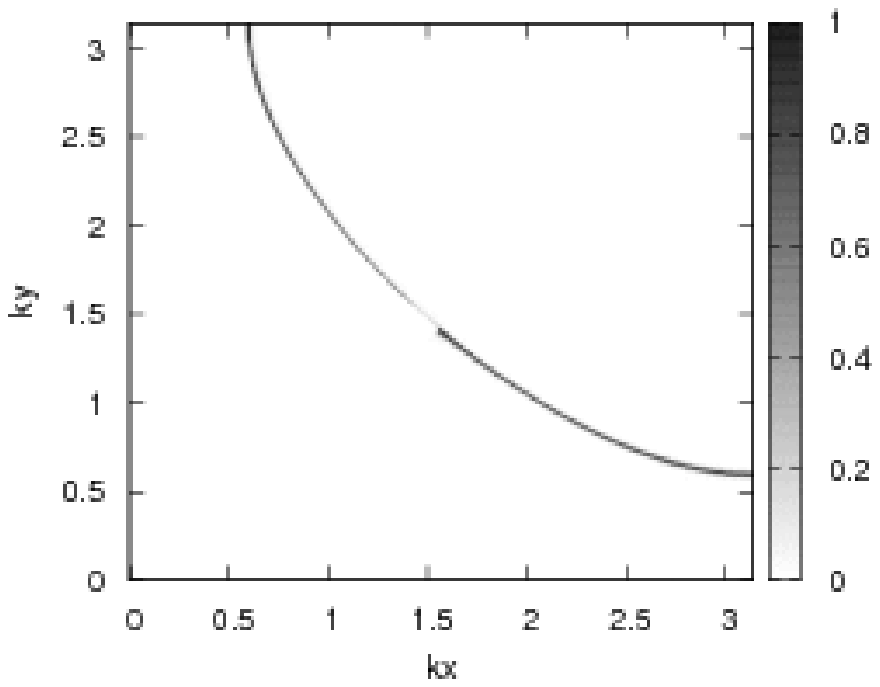}
\includegraphics[width=8cm,clip=true]{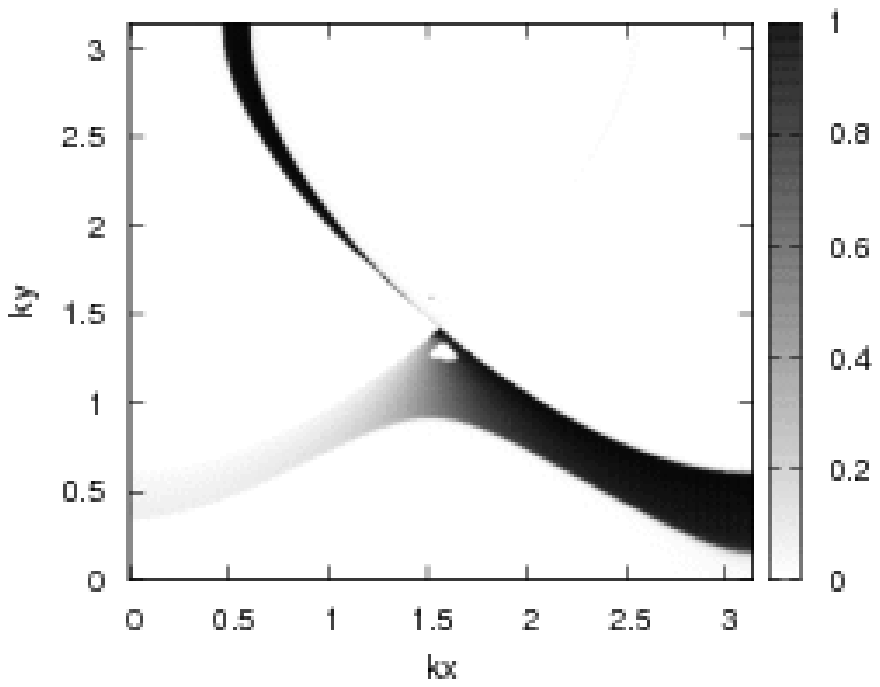}
\caption{Two dimensional FS obtained from the momentum distribution function $n_\kvec$ in Eq. (\ref{nk}).
The spectral function has been integrated over: (a) a low-energy range $W_l=10$ meV; (b) a high-energy 
range $W_l=100$ meV.}
\label{Fig10-DFS}
\end{figure}

 Figs. \ref{Fig10-DFS} (a) and (b) report the two-dimensional FS
obtained by integrating the spectral density in Eq. (\ref{nk}) over a small ($W_l=10$ meV) 
and a large ($W_h=100$ meV) energy range respectively. 
Only two poles have been considered in the recursive Green's function and the QP-mode coupling
is dressed via the 'self-consistent' vertex corrections [i.e. corrected by the 
interaction itself, see the remark after Eq. (\ref{renvertex})]. 
These corrections
are required here because at the typical fillings
we consider, the two-dimensional FS has branches connected by the critical wavevectors
(the so-called ``hot spots'' in the framework of superconducting cuprates).
The states connected by $\Qvec$ would display a gap opening similar to 
the (spurious)
one obtained in one dimension at $n=1$. For this reason we decided to 
phenomenologically suppress
this exceedingly strong effect which we attribute to the simple perturbative
treatment of the singular interactions in the model.

In the case of integration over the 
low-energy states only, one obtains a FS closely resembling 
the one of unperturbed 
free QPs'. On the contrary, upon integrating over a broad energy range, the ``FS'' appears
folded and closely tracks the one expected for a system with a static long-ranged broken
symmetry. Therefore, also in two dimensions we see that coupling the 
QPs with {\it dynamical}
modes preserves the QPs from a strong rearrangement of all the states 
(including those 
near the FS). Notice that the vertex corrections are relevant in this regard 
only around the
hot spots, but do not prevent the generic formation of shadow bands or
related features appearing in $n_\kvec$.

\section{Discussion and Conclusions}
\label{sec:disc-concl}
The results of the previous sections clearly indicate that {\it dynamical} CO/SO
fluctuations, at least in a not too strong coupling regime, do not produce substantial
effects on QPs around the FS. Therefore the absence of low-energy signatures of
CO/SO is compatible with a dynamical type of order. 
Formally we have used a 
Kampf-Schrieffer-type\cite{kam90} approach extended by a self-consistent
evaluation of self-energy corrections.
However, due to the lack of vertex corrections
our computations are restricted to weak coupling 
therefore QP weights are
close to one and the incoherent weights are small.  To have the reverse
ratios of weight we need to go to strong coupling, which is
unfortunately not feasible. Therefore we can only speculate on how the
spectral function will look like. 

We expect that
as the coupling is increased some features of our computations will persist. For example the
fact that the momentum dependent incoherent part of the spectral function is separated by
a bosonic excitation from the Fermi level is expected to be quite robust in
strong coupling. Indeed one can construct approximated excitations as in
Table~\ref{tab:excit} but with heavily dressed QPs rather
than with particles which suggest that a similar physics will be at
play. More complicated excitations may give spectral weight closer to
the Fermi level but for those the momentum dependence will be
completely washed out  providing a structureless background. 

The fact that the incoherent part should resemble the electronic
structure of the ordered state follows from continuity arguments. For example,
as one crosses a transition where the spin gets disordered at low
energies one does not expect dramatic changes in the overall
distributions of weight in the spectral
functions. Those will be determined to a large extent by the short
range correlations which may change very little across the order
disorder transition. In this sense a faint QP peak emerging
at low energies, which is certainly a dramatic perturbation from the
point of view of the Fermi liquid properties becomes a 
 small perturbation from the point of view of the spectral weight distribution.  Thus we expect this
resemblance to persist in the strong coupling limit.

The scenario we propose is of obvious pertinence in cuprates, where
standard ARPES 
experiments usually report a Fermi-liquid Fermi surface or at least
well defined nodal particles which at first sight are incompatible
with vertical stripes. On the other hand 
other experiments display the signatures
of dynamic order like the famous hourglass dispersion relation
observed in cuprates\cite{tra04} even when CO/SO is not
detected.\cite{hay04,chr04} This hourglass dispersion has been
explained computing the fluctuations on top of an ordered ground 
state which will not display nodal quasiparticles.\cite{sei05} 
Our analysis reconcile, at least qualitatively, this apparent
contradiction: fluctuating order can coexist with a Fermi-liquid like Fermi
surface (including nodal states) and still behave at energies larger
than $\omega_0$ as a stripe ordered state.

It is also interesting that recent
scanning tunneling microscopy data \cite{kohsaka} reveal the simultaneous 
existence of low energy
(Bogoliubov) QPs and high energy excitations which break 
translational invariance which thus closely agrees with the prediction
of our analysis. In particular, Fig. \ref{Fig9-DFS} clearly shows the
simultaneous 
presence of a  superconducting gap with Bogoliubov QPs at low-energy and of
high-energy spectral features, which clearly carry information on a spatial
order. For these high-energy states the construction of a local density of
states naturally displays the structure of the staggered order related to the
scattering wavevector $Q$.

As we mention in the introduction Zhou\cite{zho99,zho01} and collaborators find
different Fermi surfaces when analyzing the spectral function
integrated in a narrow window around the Fermi level or in a large
energy window. Fig.~\ref{Fig10-DFS} illustrates the weak coupling
version of these effect with the small window Fermi surface being
Fermi liquid like and the large window 
pseudo-Fermi surface showing the effects of fluctuating
order.  

The dichotomy between low and high energy spectral features as discussed
in the present paper is closely related to another dichotomy observed
in ARPES experiments, namely the one between nodal and antinodal 
QPs in underdoped cuprates.~\cite{zhou04} 
For stripe-like charge order parameter fluctuations the characteristic
scattering vectors are of a comparable magnitude than the separation
of the Fermi surface segments at the antinodal regions. As a consequence
(and quantitatively analyzed in Ref. \onlinecite{sei01b} within a
similar formalism) the scattering is much more effective for antinodal
states than for nodal ones since nodal quasiparticles get protected
by momentum conservation in addition to the dynamical protection
discussed here. Antinodal states will be affected by the vertex
corrections for which we have only a weak coupling guess and therefore
require further study but it is conceivable that spectral weight at
the Fermi level will be suppressed in these regions due to transfer of
spectral weight to the shadows.

The ascription of the dichotomy between nodal and 
antinodal QPs to low energy charge order parameter fluctuations
is also discussed in Ref. \onlinecite{sei00} within a frustrated phase 
separation scenario. For charge scattering along the Cu-O bond direction 
one also obtains a FS where the spectral
weight is confined to the nodal regions and thus is compatible with a
momentum dependent pseudogap in the sense of an anisotropy in $n_k$
along the FS.

One difficulty in interpreting data in the cuprates is the presence of
disorder. It is possible that the system has an ordered ground state
but, due to quenched disorder or a complex energy landscape\cite{sch00},
it never becomes ordered. As for a structural glass, the system has long range in time
positional order but a structural factor characteristic of a
disordered state. In this case scattering probes will fail to detect order
even though the charge is not fluctuating and is inhomogeneous. Local probes, however, should
be able to detect the ordering but those are more scarce and more
difficult to interpret. Both static disorder and our picture will
predict similar results at high energies but will strongly differ at
low energies were only within our picture one recovers a uniform Fermi
liquid. It is not clear at the moment which effect prevails in
different systems.

There are cases in which CO has been detected\cite{tra95,abb05,koh07} and still the
stripe Fermi surface computed in LDA\cite{ani04} and measured from the
$n_\kvec$ does not show up close to the Fermi level. For example no 
visible shadow FS's appears in Ref.~\onlinecite{val06},
where the FS of a CO LBCO sample only displayed Fermi arcs due to a
particle-particle pseudogap.  It is possible that in this case the
system charge orders because the commensuration with the
underlying lattice provides a pinning potential, which helps
to stabilize the charge but the spin is still quantum
disordered. Indeed spin order breaks a continuous symmetry 
whereas commensurate CO, as usually observed in cuprates, only breaks a
discrete symmetry. Therefore the former is much more fragile than the
latter. Thus we propose a state in which the charge is ordered but for
the spin our dynamical picture applies. One can still wonder 
why one does not see shadows due to the CO. One should keep in mind
that the CO is a minor perturbation to the electrons compared to SO.
In a SO state the effective potential seen by the electrons oscillates 
on a scale of $U$. In order to check the effect of CO alone without SO 
we have computed the FS in the presence of a charge modulation similar
to the one observed. The result is a large Fermi liquid like FS with
weak distortions\cite{dic07}. These may be related to features that
only very recently have  been resolved for stripe-ordered
LSCO codoped with Eu.~\cite{zabo08}   
It is only at moderate energies, above $\omega_0$ for the magnetic
excitations,  that the mixed CO-SO
fluctuations produce shadow features in the spectra and give rise to
``crossed FS's'' like those 
observed in LSCO in Refs.~\onlinecite{zho99,zho01}.

A momentum dependence of the so-called ``hump'' features at relatively high energies (of order 0.1 eV) 
has also been detected in ${\rm Bi_2Sr_2CaCu_2O_{8+\delta}}$
\cite{campuzano99}. One can speculate that scattering with a magnetic
mode related to SO as described here is responsible for this
effect. (The authors offer a related explanation involving the  
($\pi,\pi$) spin resonance). An analysis of
$n_\kvec$ similar to the one carried out for
LSCO in Refs.~\onlinecite{zho99,zho01}
could also help in discerning whether this is just a remnant effect of
the proximity 
to the insulating antiferromagnet or due to fluctuating stripes.

In conclusion, in the light of the analysis carried out in this paper, we propose that cuprates are
affected by finite-energy spin/charge fluctuations related to
proximity to ordered phases.  Whether 
CO is actually realized in a static way is rather immaterial from the 
point of view of low-energy
ARPES spectra due to the weakness of the charge modulations. 
Only when the energy is above the spin and charge fluctuating scale
 the electronic spectrum is sizably affected.
At these energies, however, the detection of CO-SO-related pseudogaps (not tied to the FS) is
quite hard due to the largely incoherent character of the spectral lines. 
Only passing through $n_k$ this dynamical tendency to order becomes visible.

We acknowledge interesting discussions with C. Castellani, C. Di Castro.
 M. G.  acknowledges financial support from MIUR Cofin 2005 n. 2005022492
and M.G and G. S. from the Vigoni foundation.

 \begin{appendix}
\section{Iterative solution of the model}
In this Appendix we give a detailed description of the iterative solution of the
two coupled equations (\ref{SELF},\ref{GF}). Inserting the lowest-order 
self-energy $\Sigma^{(1)}(k,i\omega)$ (Eq. (\ref{self0})) into 
Eq. (\ref{GF}) leads to a 
$G^{(1)}(k,i\omega)$ which displays two poles and which can be used 
to compute $\Sigma^{(2)}(k+Q,i\omega)$ (and thus $G^{(2)}(k+Q,i\omega)$)
and so on.  
This procedure therefore generates the series 
\begin{eqnarray*} 
&&G^{(1)}(k,i\omega) \to G^{(2)}(k+Q,i\omega) \to G^{(3)}(k,i\omega) \\ 
&\to& G^{(4)}(k+Q,i\omega)\to G^{(5)}(k,i\omega) \dots \to G^{(n-1)}(k+Q,i\omega)  
\end{eqnarray*} 
where $G^{(n-1)}(k+Q,i\omega)$ has a $n$-pole structure that can be 
represented as 
\begin{equation} 
G^{(n-1)}(k+Q,i\omega)= \sum_{s=1}^n \frac{\lbrack\alpha^{(n)}_s(k)\rbrack^2} 
{i\omega-E^{(n)}_s(k)}. 
\end{equation} 
Consequently the $n$-th order for the self-energy is given by 
\begin{equation} 
\Sigma^{(n)}(k,i\omega)= g^2\sum_{s=1}^n  
\frac{\lbrack\alpha^{(n)}_s(k)\rbrack^2}{i\omega \pm \omega_0 
-E^{(n)}_s(k)} 
\end{equation} 
and the sign of $\omega_0$ in the denominator depends on whether 
$f(E^{(n)}_s(k))=0,1$ (cf. Eq. (\ref{self0})). 
 
Examining the equation for the $n$-order Green's function  
$(i\omega -\varepsilon_k - \Sigma^{(n+1)})G^{(n+1)}=1$ it turns 
out that the solution can be conveniently obtained  
by solving the matrix equation: 
\begin{equation} 
\left\lbrack i\omega\underline{\underline{1}} - \underline{\underline{M}}\right 
\rbrack \underline{\underline{\cal K}} = \underline{\underline{1}} 
\end{equation} 
with 
\begin{widetext} 
\begin{equation} 
\underline{\underline{M}}= \left( \begin{array}{ccccc} 
\varepsilon_k & 
g \alpha^{(n)}_1(k) & 
g \alpha^{(n)}_2(k) & 
\cdots & 
g \alpha^{(n)}_n(k) \\ 
g \alpha^{(n)}_1(k) & 
E^{(n)}_1(k)\mp \omega_0 & 
0 & 
\cdots & 
0 \\ 
g \alpha^{(n)}_2(k) & 
0 & 
E^{(n)}_2(k)\mp \omega_0 & 
\cdots & 
0 \\ 
\vdots &\vdots &\vdots &\vdots &\vdots \\ 
\multicolumn{5}{c}\dotfill\\ 
\vdots &\vdots &\vdots &\vdots &\vdots \\ 
g \alpha^{(n)}_n(k) & 
0 & 
0 & 
\cdots & 
E^{(n)}_n(k)\mp \omega_0 
\end{array}\right) . \label{mmat} 
\end{equation} 
\end {widetext} 
and we can identify the $(n)$-order Green's function as the $(11)$-element of the 
matrix ${\cal K}$. 
Denoting by $\underline{\underline{T}}$ the transformation which diagonalizes 
$\underline{\underline{M}}$, thereby yielding $(n+1)$ eigenvalues  
$E^{(n+1)}_s(k)$,  the $(n)$-order Green's function is thus obtained as 
\begin{equation} 
G^{(n)}(k,i\omega)= \sum_{s=1}^{n+1}\frac{\lbrack T_{1s} \rbrack^2}{i\omega -E^{(n+1)}_s(k)}  
\end{equation} 
which also yields the new weights $\alpha^{(n+1)}_s(k) \equiv T_{1s}$. 
 
This scheme allows for a systematic evaluation of the Green's function up to some given 
order $(n)$. In the first step, the matrix 
$\underline{\underline{M}}$ is of order $2\times 2$ with $\alpha^{(1)}_1(k)=1$ 
(cf. Eq. (\ref{SELF})) and $E^{(1)}_n(k)=\varepsilon_{k+Q}$. 
Diagonalization yields the weights and energies of $G^{(1)}(k,i\omega)$ 
which can be used to construct the matrix for $G^{(2)}(k+Q,i\omega)$ 
(where the $(11)$-element of $\underline{\underline{M}}$ in Eq. (\ref{mmat}) 
is $\varepsilon_{k+Q}$) and so on.  
This procedure creates spectral functions with more and more poles, however, 
in case of a not too strong coupling, it will converge in the sense that the 
weight of newly created poles becomes smaller and smaller so that the 
series can be cut at the desired accuracy. 
\end{appendix}

\bibliographystyle{prsty_no_etal}

\end{document}